\documentclass[prstab,twocolumn,preprintnumbers,floatfix,nofootinbib]{revtex4}
\usepackage{amsmath}
\usepackage{txfonts}
\usepackage{textcomp,gensymb}
\usepackage{units}
\usepackage[pdftex]{graphicx}
\usepackage[pdftex]{color}
\usepackage[amsmath]{benc}
\usepackage{amssymb}
\begin{document}

\title{Three-dimensional dielectric photonic crystal structures for laser-driven acceleration}
\thanks{Work supported by Department of Energy contracts DE-AC02-76SF00515 (SLAC) and
DE-FG03-97ER41043-II (LEAP).}

\author{Benjamin M. Cowan}
\email{benc@txcorp.com}
\affiliation{Tech-X Corporation, 5621 Arapahoe Ave., Boulder, Colorado 80303, USA}
\affiliation{Stanford Linear Accelerator Center, 2575 Sand Hill Road, Menlo Park, California 94025, USA}

\begin{abstract}

We present the design and simulation of a three-dimensional photonic crystal waveguide for linear laser-driven acceleration in vacuum.  The structure confines a synchronous speed-of-light accelerating mode in both transverse dimensions.  We report the properties of this mode, including sustainable gradient and optical-to-beam efficiency.  We present a novel method for confining a particle beam using optical fields as focusing elements.  This technique, combined with careful structure design, is shown to have a large dynamic aperture and minimal emittance growth, even over millions of optical wavelengths.
\begin{center}
\emph{Submitted to Physical Review Special Topics -- Accelerators and Beams}
\end{center}

\end{abstract}


\preprint{
\begin{minipage}{3in}
\begin{flushright}
SLAC--PUB--12961, AARD--494\\
November/2007%
\end{flushright}
\end{minipage}
}

\maketitle

\section{Introduction}

The extraordinary electric fields available from laser systems make laser-driven charged particle acceleration an exciting possibility.  Because of the maturity of RF accelerator technology, it is tempting to try to scale down traditional microwave accelerators to optical wavelengths.  However, this proposal encounters several obstacles in the structure design.  First, because of the significant loss of metals at optical frequencies, we wish instead to use dielectric materials for a structure.  Second, manufacturing a circular disk-loaded waveguide on such small scales poses a significant challenge.  For these reasons, we must consider structures which differ significantly from those used in conventional accelerators.

So far, macroscopic, far-field structures have been investigated experimentally as a means to exploit high laser intensities for linear particle acceleration.  In the Laser Electron Acceleration Program (LEAP), a free-space mode was used to modulate an unbunched electron beam \cite{Plettner:LEAPResults2004}.  The experiment demonstrated the expected linear scaling of energy modulation with laser electric field as well as the expected polarization dependence.  The next stage of the experimental program, to be performed at the E163 facility at SLAC, seeks to demonstrate net energy gain by first optically bunching the electron beam using an IFEL \cite{Barnes:LEAPPhase2, Sears:PAC2007E163}.  Beyond that, the desire is to demonstrate a scalable acceleration mechanism.  For that purpose, a near-field, guided-mode structure would be more suitable.

Photonic crystals \cite{Joannopoulos:Molding} provide a means of guiding a speed-of-light optical mode in an all-dielectric structure.  They have been investigated for some time for metallic RF accelerator structures because of their potential for eliminating a major source of beam break-up instability \cite{Li:PBGWakefields,Kroll:PBG90GHz}.  In the optical regime, a  synchronous mode has been shown to exist in a photonic crystal fiber \cite{Lin:FiberAccel}.  Several years ago a study was conducted of two-dimensional planar structures \cite{benc:PCA2D}.   While the study was informative, such two-dimensional planar structures are ultimately impractical because they only confine the accelerating mode in one transverse direction.  In this paper we present a three-dimensional structure which overcomes this deficiency, confining the mode in both transverse dimensions.

In Sec.~\ref{sec:Geometry} we present the structure geometry and accelerating mode.  Then in Sec.~\ref{sec:Performance} we analyze the performance of the structure in terms of accelerating gradient and optical-to-beam efficiency.  In Sec.~\ref{sec:BeamDynamics} we address single-particle beam dynamics in the structure, presenting a focusing-defocusing lattice and demonstrating stable beam propagation.

\section{Structure geometry and accelerating mode}
\label{sec:Geometry}

\begin{figure}
\begin{center}
\resizebox{8.5cm}{!}{\includegraphics{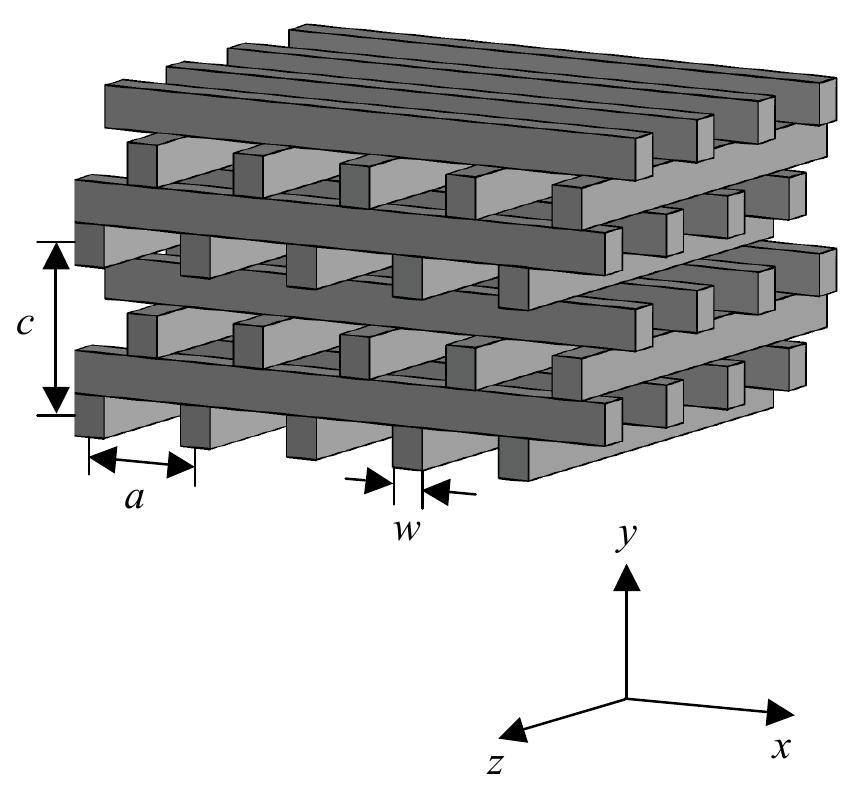}}
\caption{A diagram of 8 layers (2 vertical periods) of the woodpile lattice.}
\label{fig:woodpileLattice}
\end{center}
\end{figure}

The accelerator structure is based on the so-called ``woodpile'' geometry, a well-established three-dimensional photonic crystal lattice designed to provide a complete photonic bandgap in a structure with a straightforward fabrication process \cite{Ho:Woodpile}.  The lattice consists of layers of dielectric rods in vacuum, with the rods in each layer rotated 90\degree\ relative to the layer below and offset half a lattice period from the layer two below, as shown with the coordinate system in
Fig.~\ref{fig:woodpileLattice}.
We consider laser acceleration using a wavelength of \unit[1550]{nm}, in the telecommunications band where many promising sources exist \cite{IMRAWeb}.  At this wavelength silicon has a normalized permittivity of $\epsilon_r = \epsilon/\epsilon_0 = 12.1$ \cite{Edwards:HOCSilicon}.  We let $a$ be the horizontal lattice constant, and take $c = a\sqrt{2}$. As described in \cite{Ho:Woodpile} the lattice has a face-centered cubic structure and exhibits an omnidirectional bandgap.  Based on those results, we take the rod width to be $w = 0.28a$.  The ratio of the width of the bandgap to its center frequency is 18.7\%.

\begin{figure}
\begin{center}
\resizebox{8.5cm}{!}{\includegraphics{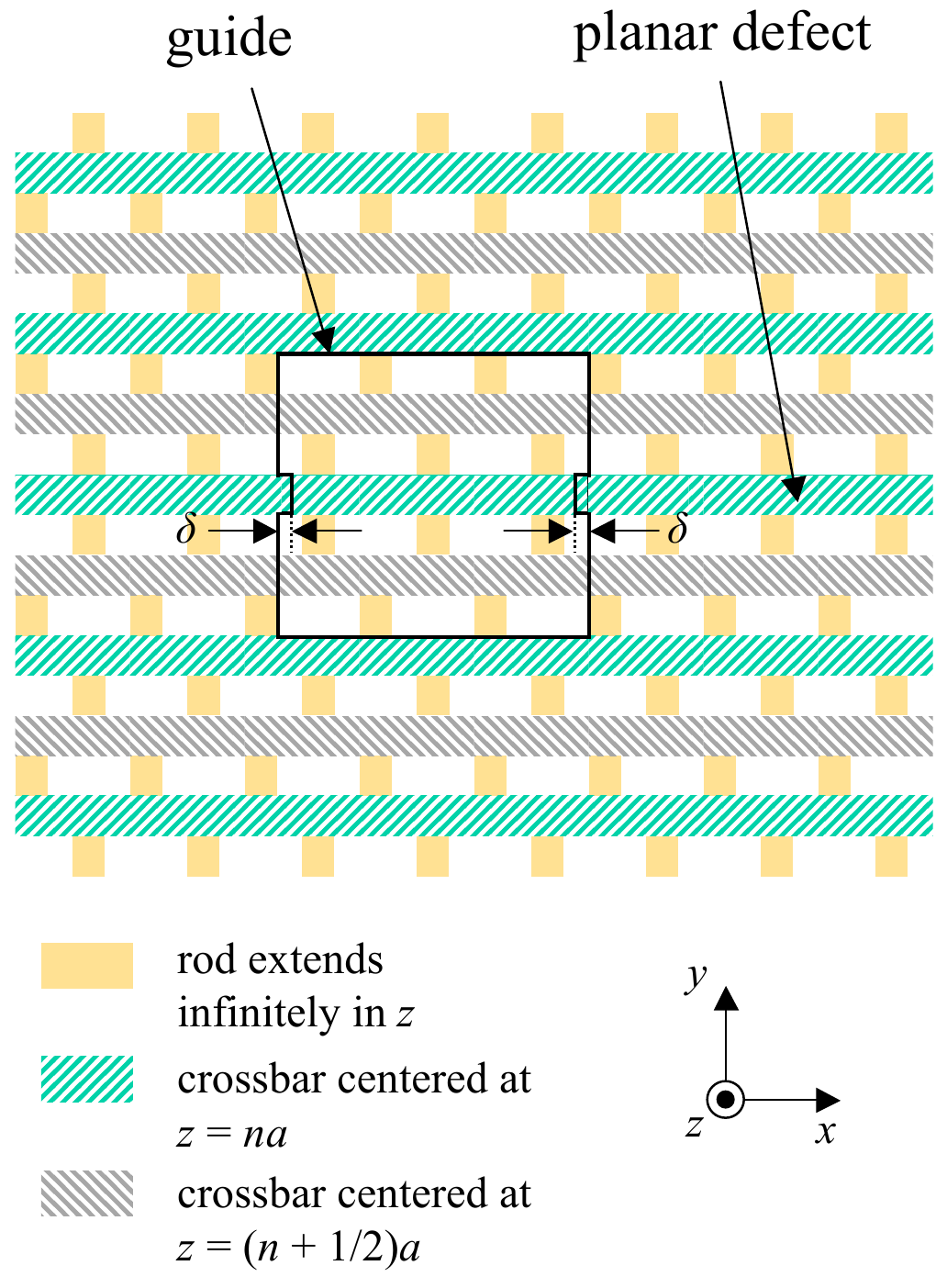}}
\caption{The geometry of a vertically symmetric waveguide structure.}
\label{fig:symmetricguide}
\end{center}
\end{figure}

\begin{figure}
\begin{center}
\resizebox{8.5cm}{!}{\includegraphics{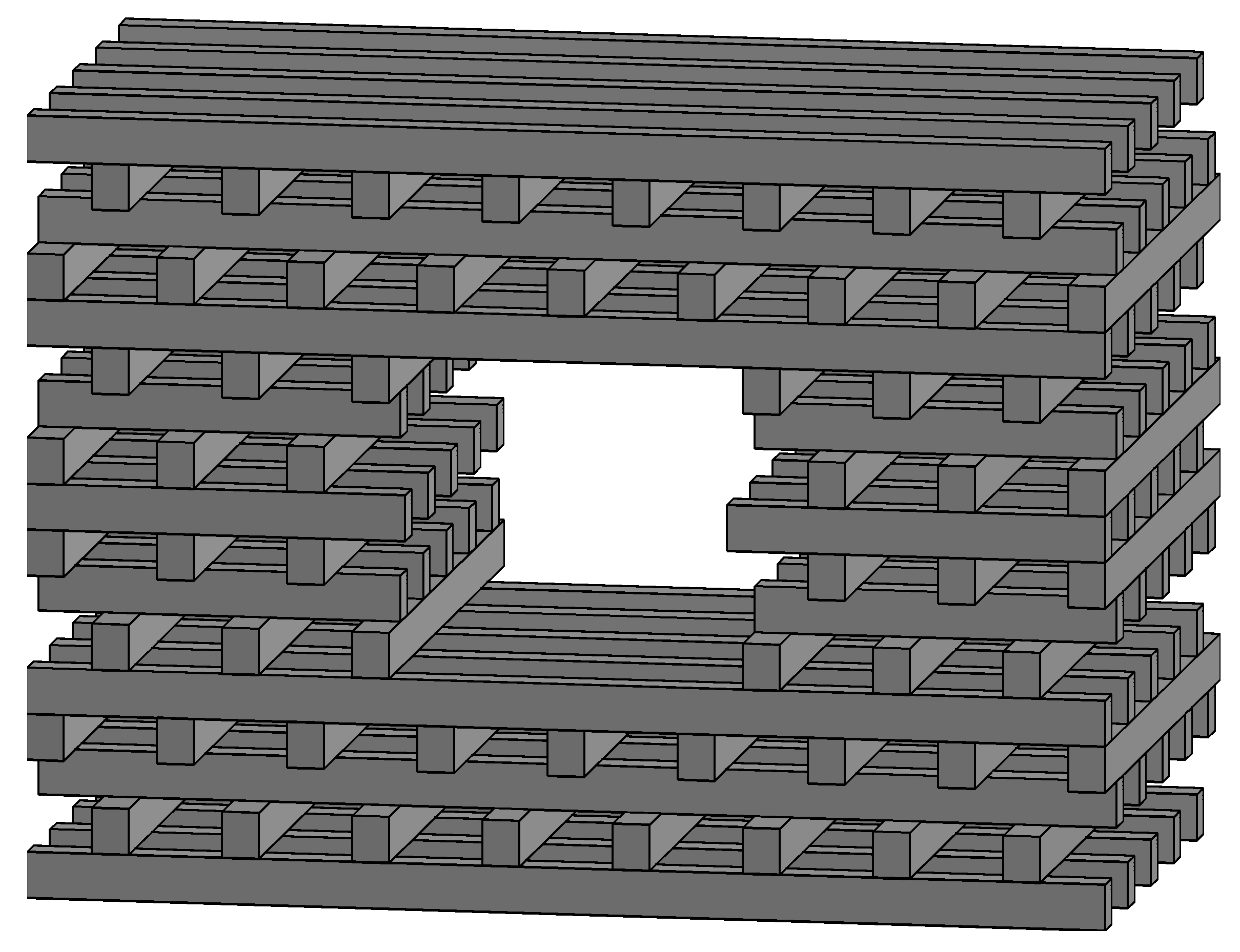}}
\caption{A symmetric waveguide.}
\label{fig:symmetricmodel}
\end{center}
\end{figure}

We form a waveguide by removing all dielectric material in a region which is rectangular in the transverse $x$ and $y$ dimensions, and extends infinitely in the $e$-beam propagation direction $z$.  In addition, there are two modifications:  In order to avoid deflecting fields in the accelerating mode, we make the structure vertically symmetric by inverting the upper half of the lattice so it is a vertical reflection of the lower half.  Second, we extend the central bars into the waveguide by $0.124a$ on each side in order to suppress quadrupole fields.  The consequences of this perturbation will be discussed further in Sec.~\ref{sec:BeamDynamics}.  The geometry, with a defect waveguide introduced, is shown schematically in Fig.~\ref{fig:symmetricguide} and visually in Fig.~\ref{fig:symmetricmodel}.

\begin{figure}
\begin{center}
\resizebox{8.5cm}{!}{\includegraphics{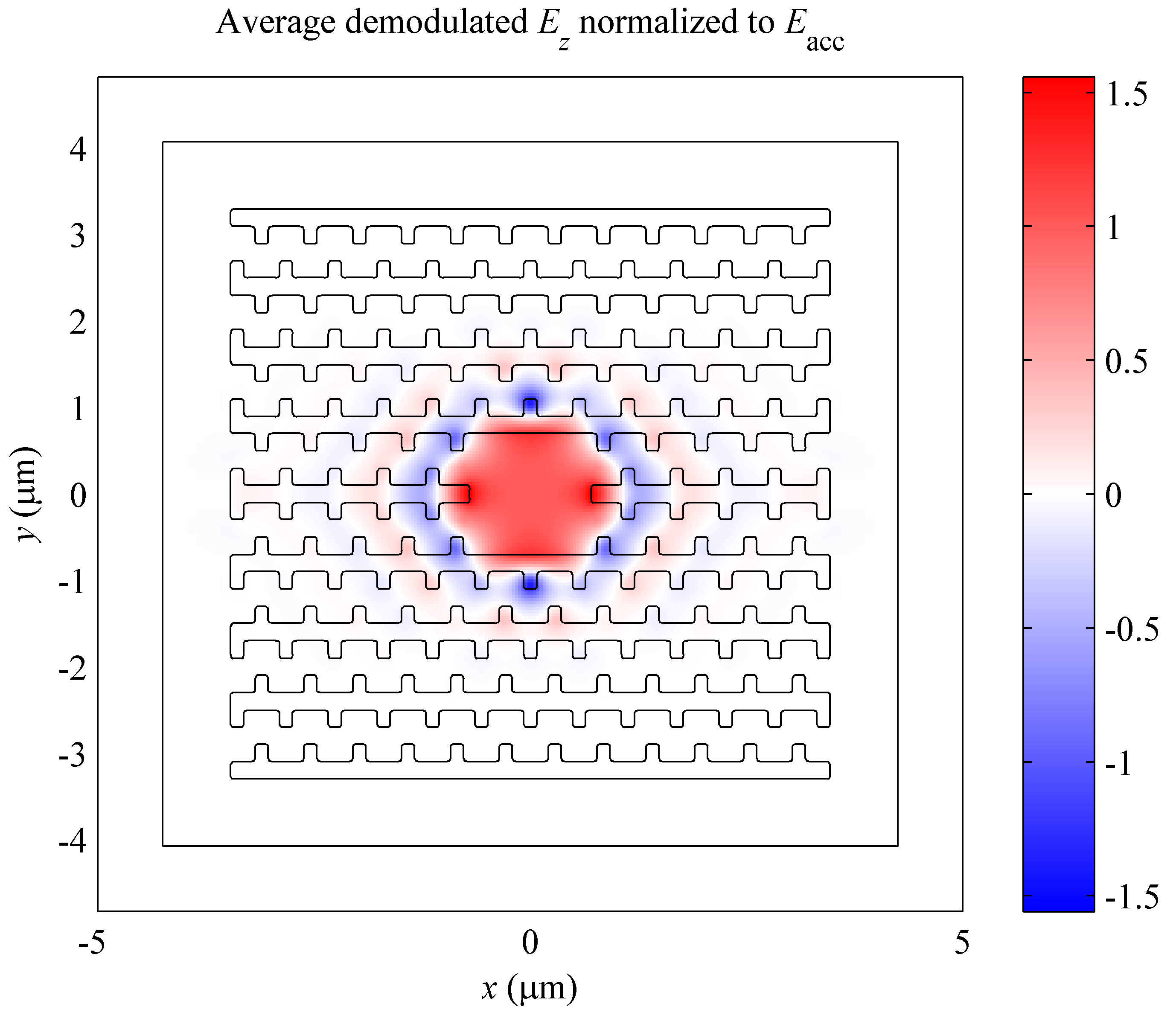}}
\caption{The accelerating field seen by a speed-of-light particle,
  averaged over a lattice period, normalized to the accelerating field
  on axis, shown with structure contours for a transverse slice at $z
  = 0$.  The inner rectangle denotes the interface between free space
  and the absorbing boundary in the simulation.}
\label{fig:SymmetricMode}
\end{center}
\end{figure}

The inversion of half the lattice introduces a planar defect where the two halves meet, but this waveguide still supports a confined accelerating mode.  Indeed, the mode is lossless to within the tolerance of the calculation, placing an upper bound on the loss of \unit[0.48]{dB/cm}.  Its fields are shown in Fig.~\ref{fig:SymmetricMode}.
For this mode $a/\lambda = 0.364$, so using a \unit[1550]{nm} source determines $a = \unit[565]{nm}$.  The individual rods are then \unit[158]{nm} wide by \unit[200]{nm} tall.  The mode was computed using the finite-difference time-domain method, including a uniaxial perfectly-matched layer to include losses \cite{Taflove:FDTD}.

\section{Performance of the accelerating mode}
\label{sec:Performance}

We can now explore the performance of this structure from the point of view of accelerating gradient and optical-to-beam efficiency.  The gradient is limited by optical breakdown to the structure material.  We can define a parameter, which is a property of the mode itself and independent of the material, which relates the accelerating gradient to the breakdown threshold of the dielectric.  While the mechanism of optical breakdown is not fully understood for most materials, let us suppose that breakdown occurs when the electromagnetic energy density exceeds a certain threshold $u\tsub{th}$ in the material.  We then define the \emph{damage impedance} of an accelerating mode by
\[ Z_d = \frac{E\tsub{acc}^2}{2u\tsub{max}c}, \]
where $E\tsub{acc}$ is the accelerating gradient on axis and $u\tsub{max}$ is the maximum electromagnetic energy density anywhere in the material.  Then we can write the sustainable accelerating gradient as
\begin{equation}
E\tsub{acc} = \sqrt{2Z_du\tsub{th}c}.
\label{eq:MaxGradient}
\end{equation}
For the mode in the woodpile waveguide, we have $Z_d = \unit[11.34]{\ohm}$.  Optical breakdown studies in silicon have shown a breakdown threshold of $u\tsub{th} = \unit[13.3]{J/cm^3}$  at $\lambda = \unit[1550]{nm}$ and \unit[1]{ps} FWHM pulse width \cite{benc:AAC2006Damage}, resulting in an unloaded accelerating gradient of \unit[301]{MV/m}.

With designs for future high-energy colliders calling for beam power exceeding \unit[10]{MW}, optical-to-beam efficiency is a critical parameter for a photonic accelerator.  Here we compute the efficiency, and the characteristics of the particle beam required to optimize the efficiency.  Unlike conventional metallic disk-loaded waveguide structures, the waveguide we consider here has a group velocity a significant fraction of the speed of light; in this case $v_g = 0.269c$.  As the particle bunch and laser pulse traverse the structure, the bunch will generate wakefields in the waveguide which destructively interfere with the incident laser pulse.  Because the beam-driven wakefield has amplitude proportional to the bunch charge, wakefields limit the practical amount of charge one can accelerate: If $q$ is the bunch charge, then the energy gained by the bunch scales as $q$, but the energy loss due to wakefields scales as $q^2$.  Increasing the charge too much ultimately decreases the energy gain, and thus the efficiency, of the structure.  To improve the efficiency, we can embed the structure within an optical cavity in order to recycle the remaining laser pulse energy for acceleration of subsequent optical bunch trains.

To compute the efficiency, we follow the treatment described in \cite{Na:Efficiency}.  According to that description, the efficiency of the structure depends upon several parameters. First, the characteristic impedance of the mode, which describes the relationship between input laser power and accelerating gradient \cite{Pierce:TWT}, is
\[ Z_c = \frac{E\tsub{acc}^2\lambda^2}{P} = \unit[484]{\ohm}, \]
where $P$ is the laser power.
Second, the group velocity of the mode affects the
efficiency, as modes with group velocity closer to the speed of light
couple better to a relativistic particle beam.  This is quantified by
the loss factor, which is given by \cite{Bane:GuideImpedance}
\[ k = \frac{1}{4}\frac{c\beta_g}{1 - \beta_g}\frac{Z_c}{\lambda^2},
\]
where $\beta_g = v_g/c$.  An optical bunch with charge $q$ will radiate fields in the accelerating mode with decelerating gradient equal to $kq$.  Finally,
the \v{C}erenkov impedance $Z_H$ parametrizes the energy loss due to
wide-band \v{C}erenkov radiation.  Following
\cite{Siemann:Efficiency}, we can estimate $Z_H$ from its value
for a bulk dielectric with a circular hole of radius $R$, which is
\[ Z_H = \frac{Z_0}{2\pi(R/\lambda)^2}, \]
where $Z_0 = \unit[376.73]{\ohm}$ is the impedance of free space.  For
arbitrary accelerating structures, we can let $R$ be a length
characterizing the radius of the vacuum waveguide, even if the guide
is not circular.  In our case the aperture is rectangular, so we
define the parameter $R$ by $R = \sqrt{A}/2 = 0.451\lambda$, where $A$
is the aperture area.  This yields $Z_H = \unit[295]{\ohm}$.

Following \cite{Na:Efficiency}, we assume the accelerator is inside an optical cavity with a round-trip loss of 5\%, and the laser pulses are coupled into the cavity through a beamsplitter with reflectivity $r$.  We consider a train of $N$ optical bunches, each
with the same charge, and assume that the duration of the train is
much less than the slippage time $\Delta\tau$ between the particles
and the laser pulse inside the structure.  We also assume that
$\Delta\tau = 3\sigma_\tau$, where $\sigma_\tau$ is the laser pulse
duration.

\begin{figure}
\begin{center}
\resizebox{8.5cm}{!}{\includegraphics{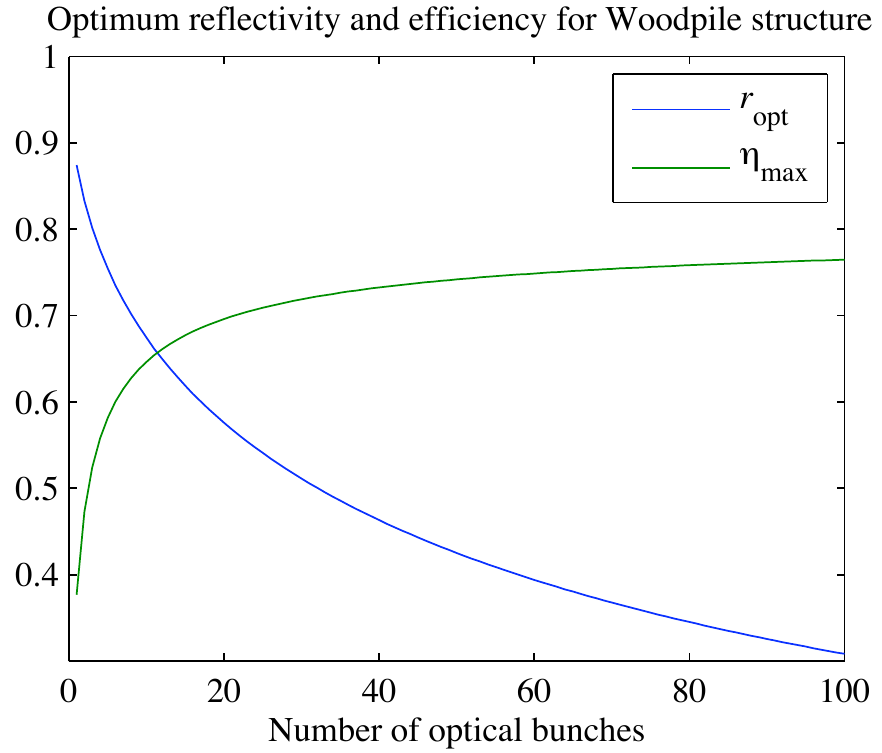}}
\caption{The optimum reflectivity and the corresponding maximum
  efficiency as a function of the number of optical microbunches in
  the bunch train.}
\label{fig:efficiency}
\end{center}
\end{figure}

\begin{figure}
\begin{center}
\resizebox{8.5cm}{!}{\includegraphics{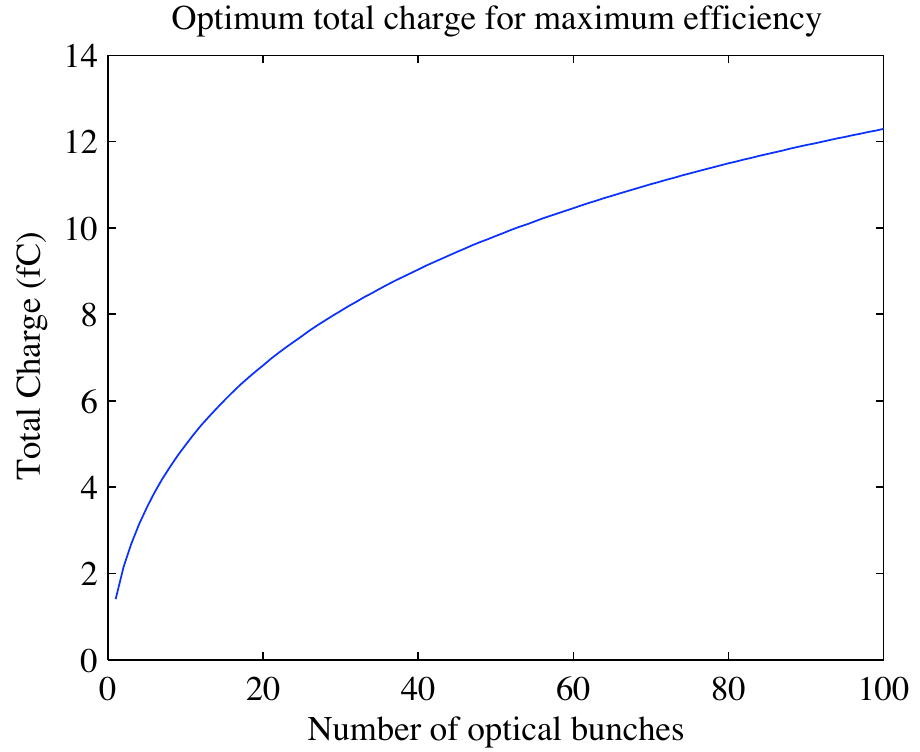}}
\caption{The optimum total charge of a bunch train as a function of
  the number of optical microbunches.}
\label{fig:charge}
\end{center}
\end{figure}

\begin{figure}
\begin{center}
\resizebox{8.5cm}{!}{\includegraphics{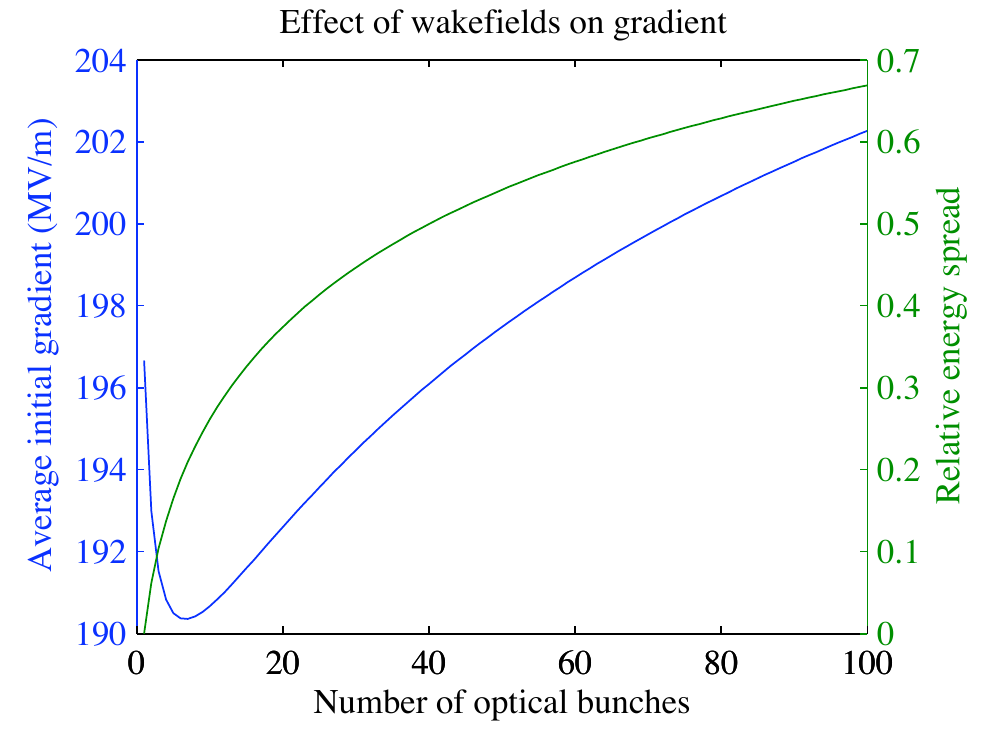}}
\caption{The average initial gradient and induced energy spread as a
  function of the number of optical microbunches.}
\label{fig:acceleration}
\end{center}
\end{figure}

As described above, trying to accelerate too much
charge can reduce the efficiency of an accelerator due to wakefield
effects.  Therefore, for a given $r$ and $N$, the structure has a
maximum efficiency $\eta\tsub{max}$.  For each $N$, we choose $r$ to
maximize $\eta\tsub{max}$.  These optimum values are plotted in
Fig.~\ref{fig:efficiency}.
From this we see that the efficiency can be made quite high.  Even
with just a single bunch, the efficiency reaches 38\%, while for a
train of 100 bunches, the efficiency is 76\%.  Once the optimum $r$ is
computed for each $N$, the total train charge $q_t$ and external laser
pulse amplitude are chosen so that they together satisfy two
conditions: First, that the peak unloaded gradient is equal to the
maximum sustainable gradient in the structure, which we take to be
\unit[301]{MV/m} from the above discussion.  Second, we require that
the charge maximize the efficiency.  We plot the optimum charge as a
function of the number of bunches in Fig.~\ref{fig:charge}.
As expected from the results in \cite{Na:Efficiency}, the
optimum total charge is low, only \unit[1.41]{fC} for a single bunch
and \unit[12.3]{fC} for 100 bunches.  With the charge having been
computed, we can then find the average initial gradient (the gradient experienced by the first optical bunch in the train) for each $N$.
We can also compute the induced energy spread on the beam as the
difference between the average accelerating gradient experienced by
the first and last bunches in the train, relative to the average
initial gradient.  These quantities are plotted in
Fig.~\ref{fig:acceleration}.
From this plot we see that wakefields have a serious effect on the
acceleration.  For just a single bunch, the average initial gradient
is reduced from \unit[218]{MV/m} (which is reduced from
\unit[301]{MV/m} due to the advancement of the electrons with respect
to the Gaussian laser pulse envelope) to \unit[197]{MV/m}.  The
minimum at around 7 optical bunches is due to two competing effects.
First, as the number of bunches increases from the single-bunch case,
the total charge increases, generating more wakefields which are
recycled within the cavity.  But we see from
Fig.~\ref{fig:efficiency} that as the number of bunches increases,
the reflectivity decreases, so a smaller fraction of these wakefields
are recycled.  More concerning is the effect on the energy spread of
the beam.  Even with only two bunches, the spread in gradient is
6.2\%, and with 5 bunches, the spread is 16.5\%.

\section{Particle beam dynamics}
\label{sec:BeamDynamics}

As remarked above, we adjusted the geometry of the structure because of beam dynamics considerations, namely to make the structure symmetric in order to
suppress deflecting fields.  Here we examine the particle beam dynamics in
this structure in more detail.  Two concerns immediately arise.
First, the structure has an extremely small aperture, approximately
$\lambda$ in each transverse dimension, so confining the beam within
the waveguide presents a serious challenge.  Second, since the
structure is not azimuthally symmetric, particles out of phase with
the accelerating field will experience transverse forces.  As it turns
out, an investigation of the second problem will yield insight into
the first:  The optical fields in the structure provide focusing
forces which are quite strong compared to conventional quadrupole
magnets and can be used to confine a beam within the waveguide.
Therefore, we begin with a description of the transverse forces in
the woodpile structure.  We then propose a focusing scheme, and
compute the requirements on the phase space extent of the particle
beam.

In our analysis of the beam dynamics in this structure, we use the
average of the fields over one longitudinal period of the photonic
crystal waveguide.  This is justified because the submicron period
together with field intensities limited by damage threshold prohibits
particles from acquiring relativistic momentum on the scale of a
single period.  To see this, one need only examine a normalized vector
potential of the fields, defined by
\[ A = \frac{eEa}{mc^2}, \]
where $E$ is the electric field magnitude.  If $A\ll 1$, then the
variation of a particle's trajectory within a period is negligible, so
the time-averaged force on the particle is well approximated by the
force due to the longitudinally averaged fields.  In our case, $E = \unit[301]{MV/m}$, so $A = 3.3\e{-4}$.  We therefore proceed to
define the longitudinally-averaged fields, as seen by a speed-of-light
particle beam.  Taking the fields to have $e^{i\omega t}$ time
dependence, we can define the averaged electric field amplitude by
\[ \bar{\vect{E}}(x, y) = \frac{1}{a}\int_0^a
\vect{E}(x, y, z)e^{ik_zz}\,dz, \]
where $\vect{E}$ is the complex electric field amplitude and $k_z = k_0 = \omega/c$ is the Bloch wavenumber in the periodic structure.  It follows that the average amplitude of the transverse force on a particle is given by
\[ \vect{F}_\perp = \frac{ie}{k_0}\del_\perp \bar{E}_z. \]
Thus $E_z$ essentially forms a ``potential'' for the transverse force.  We note that the factor of $i$ implies that particles experience a transverse force to the extent that they are out of phase with the crest of the accelerating wave.

Since the mode has speed-of-light phase velocity, $E_z$ satisfies $\nabla_\perp^2\bar{E}_z = 0$ in the vacuum region.  This implies that $\del_\perp\cdot\vect{F}_\perp = 0$, so that focusing in both
transverse directions simultaneously is impossible.  It also means that we can expand $\bar{E}_z$ as a polynomial in the complex coordinate $w = (x + iy)/\lambda$.  Using the symmetries of the structure, we find we can write $\bar{E}_z$ as
\begin{equation}
\bar{E}_z = \sum_{m=0}^{\infty} A_{2m}\re\paren{w^{2m}}.
\label{eq:EzBar}
\end{equation}
We then have that
\[ \evalu{\frac{\ptl F_x}{\ptl x}}_{x = 0, y = 0}
= \frac{ie}{\pi\lambda}A_2. \]
Thus a particle ahead in phase by $\phi$ ($kz - \omega t = \phi$) will experience focusing gradients $K_x$ and $K_y$, in the $x$ and $y$ directions respectively, given by
\begin{equation}
K_x = -K_y = -\re\paren{\frac{ieA_2}{\pi\lambda E}e^{-i\phi}}
= -\frac{eA_2}{\pi\lambda E}\sin\phi.
\label{eq:FocusingGradient}
\end{equation}
This is equivalent to a quadrupole magnet with focusing gradient $(A_2/\pi\lambda c)\sin\phi.$  The coefficients $A_{2m}$ for $m > 1$ correspond to nonlinear forces.

To examine the implications of these forces for our structure, we extract the coefficients $A_{2m}$ from the computed fields.  We do so by fitting polynomials of a form similar to that in Eq.~\ref{eq:EzBar} for $m = 0,\ldots,6$ to the computed $E_x$ and $E_y$ fields; these are sufficient to compute the coefficients of the other components using the Maxwell equations.  As noted in Sec.~\ref{sec:Geometry}, we have modified the structure geometry by extending the central bars into the waveguide in order to suppress the quadrupole fields.  This insertion distance was determined by computing the fitted $A_2$ coefficient as a function of insertion distance, and finding the value for which $A_2 = 0$.   This prevents particles from being driven out of the waveguide by the linear focusing force, but does not yet provide a mechanism for stable propagation within such a small waveguide.

\begin{figure}
\begin{center}
\resizebox{8.5cm}{!}{\includegraphics{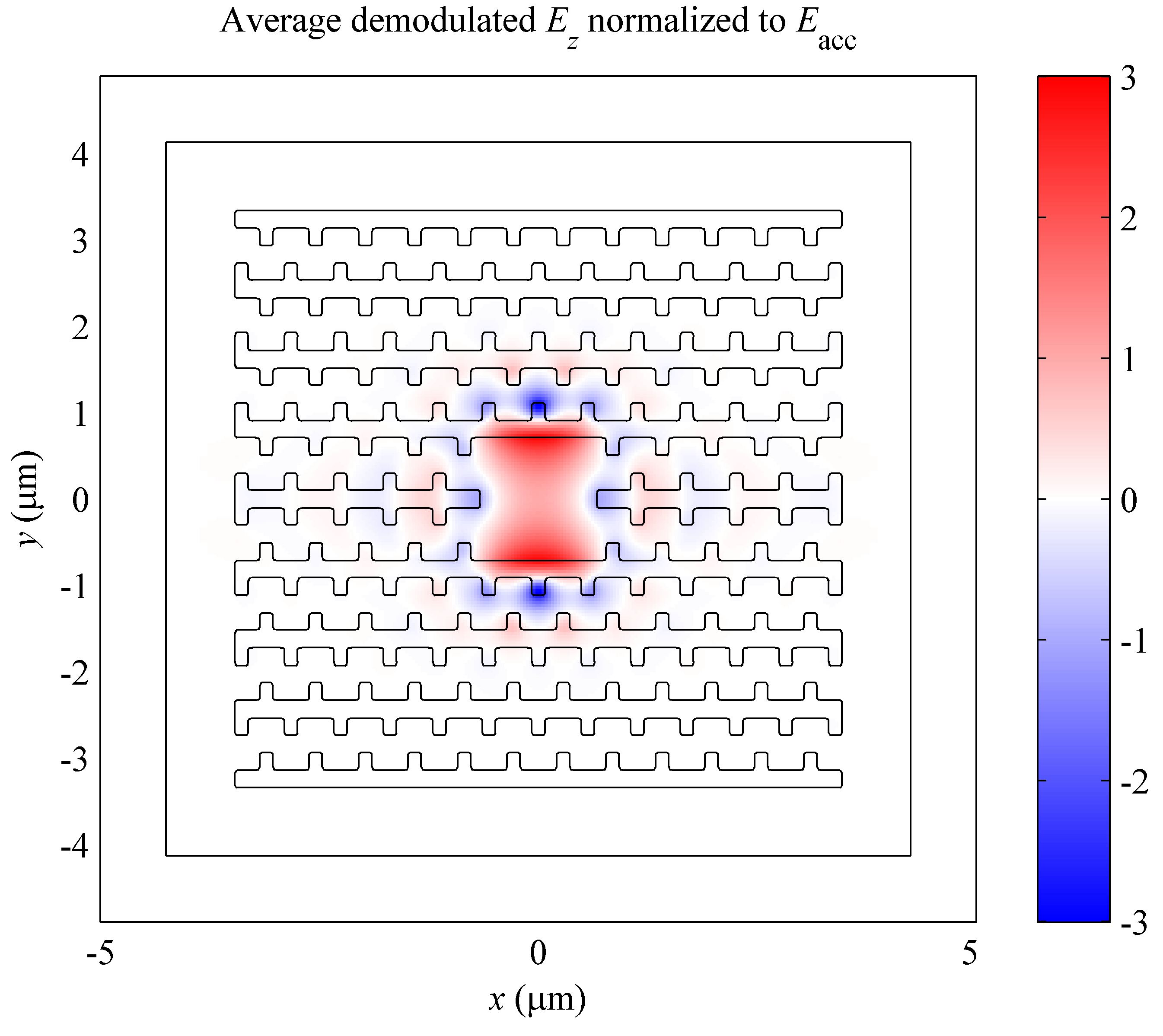}}
\caption{The accelerating field in the focusing structure modified to
  suppress the octupole moment.}
\label{fig:FocusMode}
\end{center}
\end{figure}

To address the question of stable propagation, we note from Eq.~\ref{eq:FocusingGradient} that because of the optical scale of $\lambda$, the focusing gradient can be quite large.  This suggests using these strong focusing forces to confine the particle beam by running a similar structure with the laser field $\pi/2$ out of phase with the particle beam.  We explore this possibility by again modifying our structure by inserting the central bars, this time to optimize the fields for focusing.  For this purpose we wish to suppress the lowest order nonlinear field component, in this case the octupole field given by the $A_4$ coefficient, for maximum stability.  We find that by inserting the central bars by $0.188a$, we can suppress the octupole field, while the focusing gradient for fields at damage threshold is equivalent to an \unit[831]{kT/m} magnet.  The accelerating field for this structure geometry is shown in Fig.~\ref{fig:FocusMode}.

Having found suitable optical modes for both acceleration and
focusing, we can
now describe a system of focusing elements designed to confine the
particle beam.  We consider a focusing-defocusing (F0D0)
lattice\footnote{For the remainder of this section, we refer to the
F0D0 lattice as simply the ``lattice,'' not to be confused with the
photonic crystal lattice.}, in which each cell consists of a focusing
segment of length $L_f$ run $\pi/2$ behind in phase for focusing in
the $x$ direction, followed by an accelerating segment of length
$L_a$, a focusing segment of length $L_f$ run $-\pi/2$ behind in phase
for defocusing in the $x$ direction, and then another accelerating
segment of length $L_a$ \cite{Lee:AcceleratorPhysicsF0D0}.  We are
free to choose the parameters $L_f$ and $L_a$.  Let $K$ be the peak
focusing gradient for the focusing segment, given by Eq.~\ref{eq:FocusingGradient} with
$E$ the energy of the ideal particle.  Assuming an initial ideal particle energy of \unit[1]{GeV}, we find that if we run the focusing structure at damage threshold (with an on-axis
accelerating gradient of \unit[133]{MV/m}), we have $K =
\unit[2.49\e{5}]{m^{-2}}$.  For our lattice we choose $L_f =
0.2/\sqrt{K} = \unit[401]{\micro m}$, which gives a focal length of $f =
\unit[10.0]{mm}$.  We also choose the betatron phase advance per half-cell to be $\pi/4$; then $L_a =
\sqrt{2}f = \unit[14.2]{mm}$.  The maximum beta function value is then
$\beta_F = \unit[48.4]{mm}$, the length of the unit cell is $L_c =
2(L_a + L_f) = \unit[29.1]{mm}$, and the betatron period is $4L_c =
\unit[117]{mm}$.  As the particle energy increases, we consider these lengths to scale as $1/\sqrt{K}\propto\sqrt{E}$.

The lattice described in the previous paragraph is stable under linear
betatron motion.  However, nonlinear transverse forces in the
accelerating and focusing modes may cause instabilities which limit
the dynamic aperture of the lattice and constrain the allowable
emittance of the particle beam.  To determine the effect of these
forces and compute the required emittance of a particle beam, we
perform full simulations of particle trajectories.  These simulations
include the dynamics of the particles in all six phase-space
coordinates, with field values taken from the polynomial fits.

\begin{figure}
\begin{center}
\resizebox{8.5cm}{!}{\includegraphics{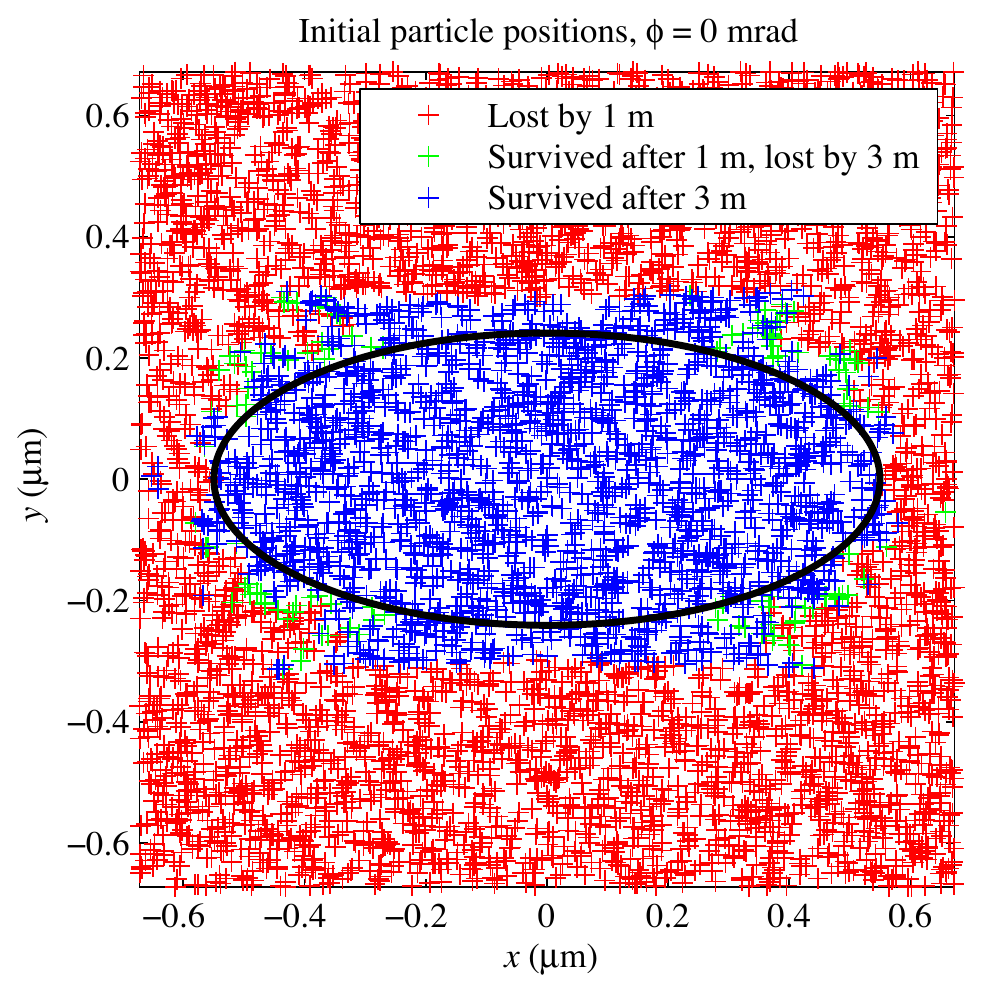}}
\caption{Initial particle positions, with colors indicating whether,
  and when, each particle exited the waveguide aperture.  The bounds of the plot correspond to the physical aperture of the structure.}
\label{fig:Aperture0}
\end{center}
\end{figure}

To compute the dynamic aperture of the lattice, we simulate particles
with initial positions uniformly distributed throughout the waveguide
aperture.  The initial transverse momenta of the particles are taken
to be 0, as are the initial deviations of the particle energies from
the ideal case.  We first simulate particles which are exactly on
crest initially.  We propagate the particles for \unit[3]{m} through
the lattice and record whether or not each particle collides with the
waveguide edge.  The initial particle positions, color-coded as to
the particle's survival, is shown in Fig.~\ref{fig:Aperture0}.
In that figure, a red marker indicates that a particle initially at
that position exited the waveguide aperture within \unit[1]{m} of
propagation.  A green marker indicates that it remained within the
waveguide through \unit[1]{m}, but exited before \unit[3]{m}, while a
blue marker indicates that the particle remained in the waveguide
through all \unit[3]{m} of propagation.  We see from the plot that
most particles that are driven out of the waveguide exited within the
first meter.  The figure also indicates that the dynamic aperture extends to almost the entire physical aperture of the structure (the dynamic aperture being wider in $x$ than $y$ because the initial longitudinal particle positions are at the high-$\beta_x$ point of the lattice).  The initial energy of the particles is \unit[1]{GeV},
and after \unit[3]{m} the energy of the ideal particle is
\unit[1.87]{GeV}.  The ellipse indicated in the figure is the ellipse
of largest area which still contains only particles which remain in
the waveguide.  If we take this ellipse to be the $3\sigma$ boundary
of the particle distribution, we can use the relation $\sigma^2 =
\beta\varepsilon$ to obtain the emittance requirement of the lattice.
We find normalized emittances of
\begin{equation}
\varepsilon^{(n)}_x = \unit[9.2\e{-10}]{m},\qquad
\varepsilon^{(n)}_y = \unit[1.09\e{-9}]{m}.
\label{eq:NormalizedEmittance}
\end{equation}
While these emittances are several orders of magnitude smaller than
those produced by conventional RF injectors, because of the small
bunch charge as discussed in Sec.~\ref{sec:Performance}, the
transverse brightness (charge per phase space area) required in the
optical structure is similar to that of conventional accelerators.

\begin{figure*}
\begin{center}
\includegraphics{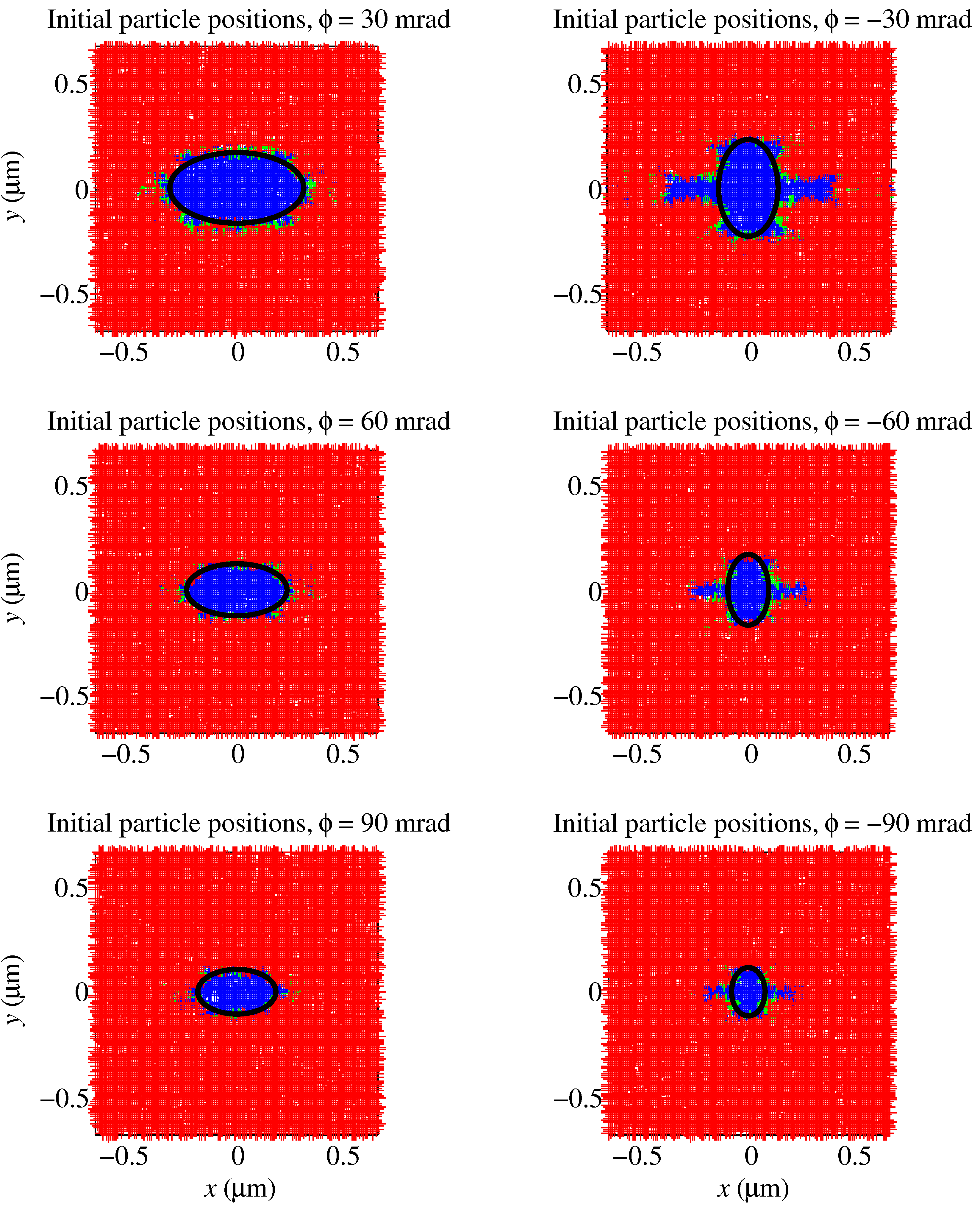}
\caption{Initial particle positions, with colors indicating whether,
  and when, each particle exited the waveguide aperture, for initial
  phases ranging from $\unit[-90]{mrad}$ to \unit[90]{mrad}.}
\label{fig:PhaseDep}
\end{center}
\end{figure*}

We can also examine the beam dynamics for particles which are slightly
out of phase with the crest of the accelerating laser field.  Running
the same simulation as above, but with the particles given an initial
phase offset, we find that the dynamic aperture is reduced as
particles become further out of phase.  This is shown in
Fig.~\ref{fig:PhaseDep}.
For particles ahead in phase ($\phi < 0$), we can see in the plots a
pattern characteristic of a fourth-order resonance, indicating that
the octupole field in the accelerating structure becomes significant
as particles become off-crest.  The difference in aperture shape
between positive and negative phase shifts could be attributed to the
sign difference of that octupole field.

\begin{figure}
\begin{center}
\resizebox{8.5cm}{!}{\includegraphics{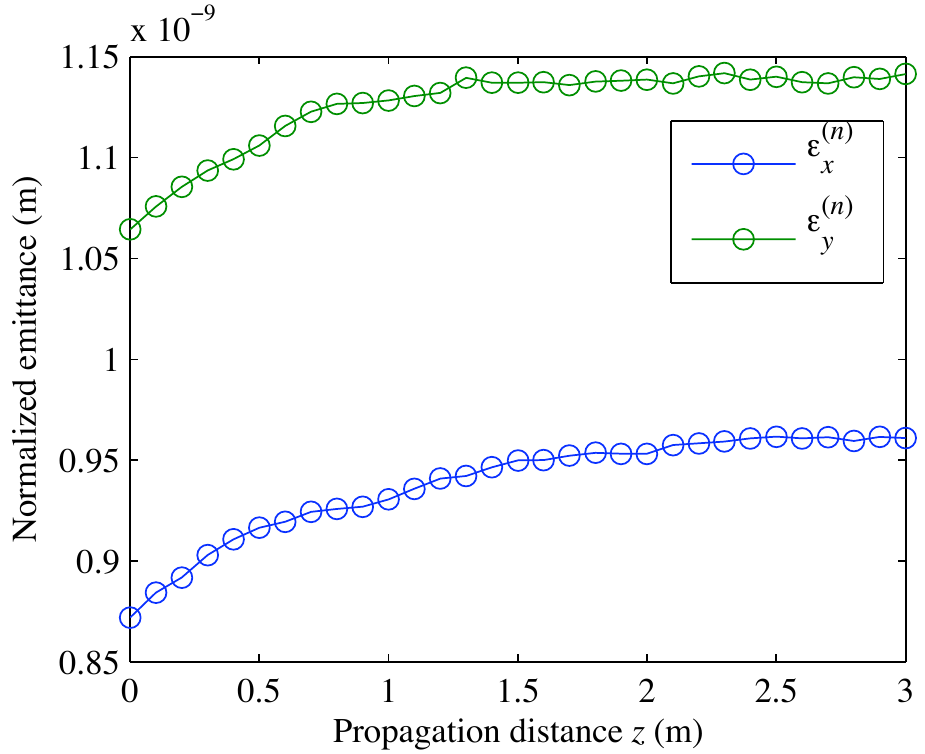}}
\caption{Transverse emittances of the particle bunch as a function of
  propagated distance.}
\label{fig:TransverseEmittance}
\end{center}
\end{figure}

With simulations of the dynamic aperture of the lattice for both on-
and off-crest particles, we can now compute the dynamics of a
realistic particle bunch, with variation in all six phase space
coordinates.  We take the initial normalized transverse emittances to be those
computed above for on-crest particles and given in Eq.~\ref{eq:NormalizedEmittance}.
We also take the RMS phase spread and relative beam energy deviation
to be $\sigma_\phi = \unit[10]{mrad}$ and $\sigma_\delta = 10^{-3}$,
respectively.  We track the ensemble of particles for \unit[3]{m} of
propagation, as before, and record the phase space coordinates every
\unit[10]{cm}.  We find that 261 of the 13228 particles simulated, or
2.0\%, exited the waveguide before completing the \unit[3]{m}
propagation.  From the recorded phase space coordinates we can track
the normalized emittances as a function of propagated distance.  The
transverse emittances are shown in Fig.~\ref{fig:TransverseEmittance}.
We see that the emittances increase initially, but that this increase
slows as the propagation continues.  For the full \unit[3]{m},
$\varepsilon^{(n)}_x$ increases by 10.1\%, while $\varepsilon^{(n)}_y$
grows by 7.3\%.  Thus the growth in invariant emittance is much
smaller than the energy gain, so that the geometric emittance of the
beam will still be adiabatically damped.  This shows that particle
bunches can be propagated stably in a photonic crystal accelerator
without an extraordinary enhancement to beam brightness.

\section{Conclusion}
We have found numerically an accelerating mode in a three-dimensional dielectric photonic crystal structure.  This mode propagates with speed-of-light phase velocity and is confined in both transverse directions with low loss, and it is therefore suitable for acceleration of relativistic particles.  Based on breakdown measurements for bulk dielectric, we estimate that the structure can sustain unloaded gradients in excess of \unit[300]{MV/m}.  We computed that the structure can operate with high optical-to-beam efficiency using the mechanism of recycling laser energy in an optical cavity.  However, the energy spread among optical bunches in a train remains a concern; some mechanism must be found to counter this effect or such an accelerator would need to operate with less than optimum charge.

We also showed a novel method for beam confinement in a small aperture which uses optical fields in a modified structure to focus the particle beam.  We found that this mechanism only required brightness comparable to conventional accelerators, and resulted in minimal emittance growth even for significant propagation distance.

Finally, while fabrication of these structures at optical lengthscales is likely to be challenging, much work on fabricating the woodpile lattice is underway in the photonics community.  Woodpile lattices with a small number of layers have been constructed using a layer-by-layer technique \cite{Fleming:WoodpileOL1999} and via wafer fusion \cite{Noda:WoodpileScience2000}.  Other experimental techniques, such as interferometric alignment \cite{Yamamoto:WoodpileAlignment} and stacking by micromanipulation \cite{Aoki:WoodpileAssembly} hold promise for efficient, cost-effective fabrication.

\begin{acknowledgments}
The author would like to thank R.~Siemann, S.~Fan, and A.~Chao for helpful advice.  Work supported by Department of Energy contracts DE-AC02-76SF00515 (SLAC) and
DE-FG03-97ER41043-II (LEAP).
\end{acknowledgments}

\bibliography{allrefs}

\end{document}